\documentclass[12pt]{iopart}
\usepackage{iopams,bbm,amsthm}

\begin{document}

\title[Security of continuous-variable QKD: exploiting symmetries in
phase space]{Security of continuous-variable quantum key distribution:
  exploiting symmetries in phase space}

\author{A Leverrier$^1$, E Karpov$^2$, P Grangier$^3$ and N J
  Cerf$^{2,4}$}

\address{$^1$ Institut Telecom / Telecom ParisTech, CNRS LTCI, 46, rue
  Barrault, 75634 Paris Cedex 13, France}
  
\address{$^2$ Quantum Information and Communication, Ecole
  Polytechnique, CP 165/59, Universit\'e Libre de Bruxelles, 50
  av. F. D. Roosevelt, B-1050 Brussels, Belgium}
 
\address{$^3$ Laboratoire Charles Fabry, Institut d'Optique, CNRS,
  Universit\'e Paris-Sud, Campus Polytechnique, RD 128, 91127
  Palaiseau Cedex, France}

\address{$^4$ M.I.T. - Research Laboratory of Electronics, Cambridge
  MA 02139, USA}
  
\ead{anthony.leverrier@enst.fr}

\begin{abstract}
  Proving the unconditional security of a quantum key distribution
  (QKD) scheme is a highly challenging task as one needs to determine 
  the most efficient attack compatible with experimental data. This task 
  is even more demanding for continuous-variable QKD as the Hilbert space
  where the protocol is described is infinite dimensional. A very natural way
  to address this problem is to make an extensive use of the
  symmetries of the protocols. In this article, we investigate a symmetry 
  in phase space that is particularly relevant to continuous-variable
  QKD, and explore the way towards a new quantum de Finetti theorem 
  that would exploit this symmetry and provide a powerful tool to assess
  the security of continuous-variable protocols.
\end{abstract}

\maketitle

\section{Introduction and motivation}

The greatest novelty brought by Quantum Key Distribution (QKD) is
that, for the first time, a secret key agreement scheme can be proven
unconditionally secure, that is, without making any assumptions about
the power of an adversary.

Even if this claim has been made repeatedly since the proposal of the
first QKD protocol in 1984 \cite{BB84}, it is only recently that
complete proofs of security have been rigourously established. Proving
the security of a scheme without making any simplifying assumptions is
indeed quite challenging: the legitimate parties, Alice and Bob, need
to infer what is the most efficient attack that an eavesdropper, Eve, could
perform. This can be achieved by considering all bipartite states
$\rho_{AB}$ compatible with Alice and Bob's data, but this quickly
becomes almost untractable since the dimension of the Hilbert space
$\mathcal{H}^{\otimes n}$ relevant to describe $\rho_{AB}$ grows
exponentially with the number $n$ of quantum signals exchanged during
the protocol. As a consequence, security proofs were often derived
while restricting the adversary to the so-called {\em
collective attacks}. In such attacks, the state $\rho_{AB}$ is
supposed to be {\em independent and identically distributed} (i.i.d.),
meaning that there exists a state
$\sigma_{AB} \in \mathcal{H}$ such that $\rho_{AB} =
\sigma_{AB}^{\otimes n}$. As a consequence, the Hilbert space needed
to analyze the protocol becomes $\mathcal{H}$ instead of
$\mathcal{H}^{\otimes n}$: no need to emphasize that this ``small''
assumption considerably simplifies the analysis !

The question then is to know whether such a hypothesis limits the
power of the adversary in a non trivial way, or, said otherwise, whether 
this leads to an unreasonably optimistic view of the security of QKD.
Fortunately, this is not the case as collective attacks were
recently proven asymptotically optimal against protocols described with
a finite-dimensional Hilbert space \cite{R07}. The main tool used to
answer this problem was a quantum de Finetti theorem which can be
roughly summarized as saying that a certain class of states in
$\mathcal{H}^{\otimes n}$, namely {\em symmetric states}, can be well
approximated by mixtures of i.i.d. states. From a cryptographic point
of view, this means that general {\em symmetric} attacks are almost the same
as collective attacks. The last step to complete the proof is to show
that a symmetric attack is optimal for the eavesdropper, or,
equivalently, that the state $\rho_{AB}$ can safely be assumed
symmetric, which is indeed the case for most QKD protocols.

The quantum de Finetti theorem is thus quite powerful as it allows to
derive the security of a QKD scheme against arbitrary attacks as soon
as its security against collective attacks is proven. Moreover, the
full security is obtained almost for free, in the sense that the
decrease of key size caused by allowing the adversary to perform any
non-collective attack is negligible, at least in an asymptotic regime. 
In a finite size scenario, however, the impact on the key size could be 
significant, although it should be compared with other finite-size effects
such as the precision of parameter estimation or the efficiency of 
error correction \cite{SR08}. In this context, alternatives to the
de Finetti theorem might also be worth investigating 
as they can lead to improved bounds \cite{CKR09}.

Unfortunately, the application of the quantum de Finetti theorem 
that was presented in \cite{R07} is restricted to QKD schemes that are
described in a finite-dimensional Hilbert space. Apart from the fact
that decribing any protocol in a finite-dimensional Hilbert space is
nothing less than an approximation (even though quite a reasonable one for
protocols involving qubits), it is clear that it does not apply to
protocols genuinely described in infinite-dimensional Hilbert spaces
such as protocols explicitly built on the continuous amplitude 
components of the light field (see \cite{CG07} and
references therein). The reason for this is that
the quantum de Finetti theorem fails as soon as the dimension $d$ of
the Hilbert space $\mathcal{H}$ is not small compared to the number
$n$ of subsystems considered, which is obviously the case if $d$ is
infinite. Moreover, not only is the present version of the quantum de
Finetti theorem limited to low dimensional Hilbert spaces, but
counter-examples have been exhibited that demonstrate that 
a dimension-independent de Finetti theorem cannot exist \cite{CKMR07}.
 
Nevertheless, the impossiblity of a general dimension-independent
theorem does not rule out the possibility of more restricted versions
of the theorem, which might still be highly relevant to prove 
the security of QKD schemes. In particular, the quantum de Finetti theorem
of \cite{R07} is concerned with 
(permutation) symmetric states in $\mathcal{H}^{\otimes n}$, that is, 
states that are invariant under arbitrary permutations of their $n$ subsystems.
The only approach that has been pursued to date in order to extend the range 
of application of this theorem to infinite-dimensional Hilbert spaces
has consisted in restricting the set of states in such a way that 
a finite-dimension theorem can be applied \cite{KM07,DOS07,RC08,KW09}. 
In particular, in \cite{RC08}, the quantum de Finetti theorem of \cite{R07} 
has been applied to infinite-dimensional quantum systems conditioned 
on certain measurement results. The main consequence is 
to rigorously justify the assumption of a finite-dimension theorem
in the context of QKD protocols using attenuated coherent pulses 
as the support for qubits. The theorem can then be used to derive 
the security of some continuous-variable schemes \cite{LG09} as long as the energy of
the signal states is not too important. The main drawback is that some 
experimental conditions need to be checked, which was not the case for
finite-dimensional protocols.

In this paper, we explore a radically different approach, which might greatly simplify the
security proofs of continuous-variable QKD. The idea is to derive a new quantum de Finetti theorem 
corresponding to symmetry classes other than permutations of the subsystems. 
Our main insight is to describe the protocol in {\em phase space} instead of Fock
representation, and to study a symmetry group that is specific to phase space. 
This choice features several advantages. First, the
phase space representation is the natural choice for the analysis of
continuous-variable QKD where the information is typically encoded
onto the quadratures of the light field (see \cite{CG07}). 
Moreover, if collective attacks are indeed
asymptotically optimal as they generally are for discrete-variable QKD, 
it would be nice to have an interpretation of
this result using covariance matrices. It should be pointed out that
when restricted to collective attacks, the security of the protocol is
completely characterized by the covariance matrix of the system shared
by Alice and Bob \cite{GC06,NGA06}. Last but
not least, the phase space representation has the remarkable property
to be finite dimensional: one just traded a discrete description in a
infinite-dimensional Hilbert space (the Fock representation) for a
{\em continuous} description in a finite-dimensional real space. 

Interestingly, in a classical setting, versions of the de Finetti theorem 
that apply to orthogonally-invariant continuous probability distributions
have been known since a long time. Here, we make the first steps 
towards the generalization of this theorem to a quantum setting. 
Obviously, since a general dimension-independent quantum de Finetti theorem
is impossible, we cannot hope to establish one just by switching between
an infinite-dimensional state-space representation and a 
finite-dimensional phase-space representation. The trick is that the symmetry hypotheses
needed for the phase-space de Finetti theorem are stronger than the ones 
used in the previous quantum versions of the de Finetti theorem. We will
show that these stronger symmetry hypotheses are, however, perfectly compatible
with continuous-variable QKD protocols.

The outline of the paper is as follows. In Section \ref{symmetry}, we
explicitely make the link between symmetry properties and security
proofs. In Section \ref{cvqkd}, we present continuous-variable
QKD and introduce a new symmetry for such protocols. This symmetry is
then presented in more details in Section \ref{rotation-invariance}, where
we make the preliminary steps towards the derivation of a new quantum de Finetti theorem
for continuous variables. Finally, the conclusions are drawn in Section \ref{conclusion}.

\section{Role of symmetry in the security proofs}
\label{symmetry}



The goal of this Section is to explain how symmetry considerations can
simplify the theoretical analysis of quantum cryptography. In
particular, we would like to provide a theoretical justification to
the common attitude of considering the state $\rho_{AB}$ shared by
Alice and Bob as being symmetric. Note that a more mathematical
argument can be found in \cite{CKR09}.

As we mentioned previously, applying a de Finetti theorem to prove
the security of QKD protocols works if one can first assume, without
loss of generality, that the state $\rho_{AB}$ is symmetric, that is
invariant under any permutation of its $n$ subsystems. Here, we show
that this assumption is justified, but that one is actually not limited
to considering the action of this particular symmetry group $\mathcal{S}_n$
(one can also consider larger symmetry groups). Basically, the idea is that by assuming any symmetry, Alice and Bob will always underestimate the secret key rate they can extract from their data.

The secret key rate for a particular instance of a QKD protocol is a
function of the state $\rho_{AB}$ shared by the legitimate parties,
Alice and Bob. The eavesdropper, Eve, is assumed to have the maximal
information compatible with $\rho_{AB}$ meaning that her state
$\rho_E$ is such that $\rho_E = \tr_{AB}(|\Psi_{ABE}\rangle)$ where
$|\Psi_{ABE}\rangle$ is any purification of $\rho_{AB}$. Note that all
purifications are equivalent up to a unitary operation applied on 
system $E$.  More precisely, $\rho_{AB}$ represents the {\em
  knowledge} that Alice and Bob have about the quantum state they
share. For this reason, $\rho_{AB}$ is subjective and inevitably
depends on assumptions made by Alice and Bob. It must be emphasized
that this cannot be avoided by performing a quantum tomography of the
state since the latter is also subject to hypotheses, namely that one has
access to an arbitrary large number of independent and identical
copies of a single state. The exponential version of the quantum
de Finetti theorem as derived in \cite{R07} gives a partial answer 
to this problem: if $\rho_{AB}$ is invariant under permutations of its subsystems,
then it can be well approximated by a mixture of i.i.d. states, so quantum
tomography is therefore justified.

A crucial observation is that Alice and Bob would like to ignore or
forget the properties of $\rho_{AB}$ they are not interested in,
typically possible correlations between the $n$ subsystems of their state,
hence obtaining $\rho_{AB} = \sigma_{AB}^{\otimes n}$ for some
prototype state $\sigma_{AB} \in \mathcal{H}$. Unfortunately, this
action of forgetting comes at a price, namely erasing some potentially useful
information. The first idea to make the argument more rigourous is
that Alice and Bob can actually {\em enforce} the symmetry they
want. Let us for instance consider symmetry under permutations of the
subsystems of $\rho_{AB}$ which is the symmetry commonly used in
various QKD security proofs (with the notable exception of protocols
such as the Differential Phase Shift (DPS) \cite{IWY02} or the
Coherent One-Way (COW) \cite{STS05}). This symmetry can be
enforced in the following way: Alice and Bob can perform the
same random permutation $\pi$ over their respective state, with $\pi$ being
chosen uniformly over the symmetric group $\mathcal{S}_n$. 
This operation transforms $\rho_{AB}$ into
\begin{displaymath}
  \frac{1}{n!} \sum_{\pi \in \mathcal{S}_n} \pi  \rho_{AB} \pi^{\dagger} \otimes |\pi\rangle\langle \pi|_C
\end{displaymath}
where $\pi$ is the unitary operator implementing the
permutation $\pi$ to both systems $A$ and $B$, $\{|\pi\rangle\}_{\pi}$ is an orthogonal family of vectors and $C$ is
a classical auxiliary space whose sole purpose is to store the
information concerning the permutation $\pi$ that was applied. Then, tracing over
system $C$ (or equivalently giving this system to Eve), Alice and Bob
obtain the state $\bar{\rho}_{AB}$, which is symmetric by construction.
Obviously, for any practical purpose, applying such a procedure is out
of question as it would at least involve a quantum memory in order to
store each subsystem while Alice and Bob wait for the total state
$\rho_{AB}$. One may object, however, that applying such a permutation $\pi$
to $\rho_{AB}$ is equivalent to merely relabeling the indices of Alice
and Bob's data, which is much simpler to implement. The key
is that both procedures are indistinguishable, which is a clear consequence 
of the fact that the permutation of subsystems commutes
with the measurement procedure and classical post-processing. This is true 
for most protocols, such as BB84 or continuous-variable protocols, 
but not for DPS or COW. In order for the two procedures 
to be completely equivalent, Alice and Bob should completely forget 
which particular permutation was performed. A second crucial point is
that, in reality, Alice and Bob do not even need to permute the labels 
of their data. What is really necessary is that they should never use 
any information related to the order of their data (the labeling of
their data) when they extract the key.

It must be realized that enforcing such a symmetry can only decrease the
secret key rate since Alice and Bob give additional information to Eve, or,
equivalently, forget some {\em a priori} available information. On the other
hand, while they are only throwing information that they do
not know use in practice (the labeling of their data), the impact of this
symmetrization step to the key rate is actually negligible. 
Note than nothing forbids one to use such a technique in the study of 
the DPS and COW protocols. However, correlations between different subsystems
are essential for these protocols to work and no key could be extracted if
one was forgetting them. In principle, any symmetrization is applicable
to any QKD protocol, but some symmetrization procedures essentially
erase all the relevant information and are consequently useless for
the study of such protocols. Other symmetries have been investigated
in the literature, for instance random bit-flip or phase-flip applied
simultaneously by Alice and Bob, and have led to simplifications in the
analysis of some protocols \cite{KGR05}.

The above reasoning can easily be generalized to other symmetries. 
Let $\mathcal{G}$ be a symmetry group in $\mathcal{H}^{\otimes n}$.
Alice and Bob can perform a random $g$ drawn from $\mathcal{G}$ 
and later forget about $g$, thus transforming
$\rho_{AB}$ into
\begin{displaymath}
  \bar{\rho}_{AB}^{\mathcal{G}}=\tr_C \left(\frac{1}{\# \mathcal{G}}
    \sum_{g \in \mathcal{G}} g\; \rho_{AB} \; g^{\dagger} \otimes |g\rangle\langle g|_C\right)
\end{displaymath}
where $\#\mathcal{G}$ is the cardinal of $\mathcal{G}$. The group $\mathcal{G}$ can even be continuous, in which case the
discrete sum should simply be replaced by an integral over the Haar
distribution of $\mathcal{G}$. This is actually what we will do for
continuous-variable protocols.


\section{Phase-space symmetry for continuous-variable QKD}
\label{cvqkd}

\subsection{Brief summary of the theoretical analysis of
  continuous-variable QKD}

We rapidly describe continuous-variable protocols, 
as a more detailed presentation can be found in \cite{CG07}.
Continuous-variable QKD comes with two flavours depending on whether the
quantum state shared by Alice and Bob is characterized by a quantum
bit error rate (QBER) or by a covariance matrix. In the first category
lie protocols where the quadratures of the light field are just the
support for encoding bits. Such protocols usually use postselection to
improve their QBER \cite{SRL02}. Then, the analysis is somewhat similar
to that of discrete-variable protocols. In the second category of
protocols such as \cite{GG02}, with which we are only concerned here,
the quantum state shared by Alice and Bob is
characterized by its covariance matrix. This means that the
continuous-variable approach is used even for the description of the
state, and not only as a means to carry information over quantum
channels. This approach has an important drawback, namely that
postselection is a priori impossible: one must keep all data. This
is particularly damaging during the classical post-processing of the
protocol where one now has to deal with real random variables instead
of binary random variables. The main implication is that the
reconciliation step, which roughly corresponds to correcting
discrepancies between Alice and Bob's classical data, becomes a task 
that is much more involved than correcting errors between two binary
strings. The error rate may indeed be much higher as the protocols 
with continuous variables may tolerate very low signal-to-noise ratio.
The classical problem of reconciliation was until recently
limiting the range of continuous-variable QKD protocols
\cite{LAB08}. However, simpler discrete modulation schemes 
can help dealing with the reconciliation problem \cite{LG09}.


\subsection{Rotation symmetries for continuous-variable QKD}


One of the nice features of continuous-variable QKD is that the security 
against collective attacks is entirely characterized by the covariance
matrice of $\rho_{AB}$. As we restrict our analysis to collective attacks,
one has $\rho_{AB} = \sigma_{AB}^{\otimes n}$, and the covariance matrix 
$\Gamma$ of $\sigma_{AB}$ is usually assumed to be of
the form:
\begin{equation}
  \Gamma_{\mathrm{sym}} =
  \left(
    \begin{array}{cc}
      X \mathbbm{1}_2 & Z \sigma_z \\
      Z \sigma_z & Y \mathbbm{1}_2 \\
    \end{array}
  \right),
\end{equation}
where $\sigma_z = \mathrm{diag}(1,-1)$. Note that this form can easily
be understood from an experimental point of view since the quantum
channel is not supposed to induce correlations between different
quadratures, for instance, but no theoretical justification has been
given so far. Here, we use the ideas explained in previous section to prove
that $\Gamma$ can indeed be supposed to take this simple form.

Since we make the assumption of a collective attack, the covariance matrix $\Gamma$ is
well defined and can be estimated by Alice and Bob. The most
general form for $\Gamma$ is:
\begin{equation}
  \Gamma =
  \left(
    \begin{array}{cccc}
      X_{11}& X_{12} & Z_{11} & Z_{12} \\
      X_{12}& X_{22} & Z_{21}& Z_{22} \\
      Z_{11}& Z_{21}& Y_{11}& Y_{12} \\
      Z_{12}& Z_{22}& Y_{12}& Y_{22}\\
    \end{array}
  \right).
\end{equation}
The idea is that Alice and Bob can perform some symmetrization operation,
which transforms $\Gamma$ into $\Gamma_{\mathrm{sym}}$. First, note that
their classical data are two strings $x, y \in \mathbb{R}^n$,
which correspond to the results of homodyne measurements of the various quadratures of $\rho_{AB}$. The
reconciliation is always optimized for a Gaussian channel, meaning
that the random variable $y$ is modelled as $y = t x + z$ \cite{LAB08}
where $t$ is a transmission factor and $z$ is a random variable
modelling the added noise and characterized by its variance
$\sigma^2$. Therefore, the reconciliation procedure would not be
affected if Alice and Bob both performed the same random orthogonal
transformation $R \in \Or(n)$ to their respective data, since one would
then have $Ry = t Rx + z'$, where $z'$ is a rotated noise with the same
variance $\sigma^2$.  If Alice and Bob apply such a random orthogonal
transformation (rotation) and forget which one has been performed, 
their data become ``symmetric'' in the sense that the matrix
$\Gamma$ takes the form of $\Gamma_{\mathrm{sym}}$ where $X =
(X_{11}+X_{22})/2, Y = (Y_{11}+Y_{22})/2$ and $Z = (Z_{11}-Z_{22})/2$.
The fact that the covariance matrix $\Gamma_{\mathrm{sym}}$ features
$Z \sigma_z$ instead of $Z \mathbbm{1}_2$ simply reflects the fact that
$\Gamma_{\mathrm{sym}}$ is not the covariance matrix of the classical
data of Alice and Bob in the {\em prepare-and-measure} scenario, 
but the covariance matrix of $\rho_{AB}$ in the equivalent 
{\em entanglement-based} scenario. In the latter case, Alice and Bob
would actually apply {\em conjugate} orthogonal transformations 
to their respective share of the state instead of the same transformation. 
By conjugate transformation, we mean the transformation whose 
corresponding $2n\times 2n$ matrix in phase space 
is obtained from the original one by flipping the sign of all rows 
whose label corresponds to a $p$ quadrature and then flipping the sign 
of all columns whose label corresponds to a $p$ quadrature. 
This can be understood by considering a two-mode squeezed vacuum, 
which is the state characterizing the inherent symmetry of 
continuous-variable QKD: this state has a covariance matrix 
$\Gamma_{\mathrm{sym}}$ where $Y=X$ and $Z = \sqrt{X^2-1}$,
and is invariant under conjugate orthogonal transformations 
performed by Alice and Bob.

As we will see, this new symmetrization (based on orthogonal 
transformations in phase space instead of permutations in state 
space) has several crucial consequences. First, it allows us
to rigorously prove that Alice and Bob can safely assume their
covariance matrix to have a simple structure,
characterized by only three parameters which are easily estimated 
experimentally (this was done until now
with no firm theoretical justification). The second consequence,
which we will study in the next Section, is that it gives a simple
structure to the state $\rho_{AB}$ which enables us to investigate
the unconditional security using a de Finetti approach.

\section{Invariant states under rotations in phase space}
\label{rotation-invariance}

The goal of this Section is to give some insights on the structure
of the states which are invariant under orthogonal transformations in phase space. 
More precisely, if Alice and Bob perform $n$ homodyne measurements on $\rho_{AB}$ 
(Alice and Bob are assumed to measure the same quadrature since they discard the data corresponding to measurements of incompatible quadratures), they obtain two random vectors $x,y \in
\mathbb{R}^n$. We are interested in unitary transformations whose
effect on $\rho_{AB}$ is described by an orthogonal transformation 
on the probability distributions of $x$ and $y$. As these
probability distributions are completely characterized 
by the Wigner function of $\rho_{AB}$, the states of interest are 
simply those whose Wigner function is invariant 
under such symplectic transformations.

Before describing the bipartite case, it is useful to consider first the
single-party case, where the state of Bob is traced out.


\subsection{Single party case}

Here, we are interested in generalizing the concept of {\em
  orthogonally invariant} probability distributions to the quantum
setting, that is, to Wigner functions.  A $n$-mode state $\rho$ is
termed {\em orthogonally invariant} in phase space if it is invariant
under the action of any $n$-mode Gaussian unitary operator
corresponding to a real symplectic orthogonal transformation 
(or, simply said, a rotation) in the
$2n$-dimensional phase space of $\rho$. Physically, this means that
$\rho$ remains unchanged after being processed via any $n$-mode
passive linear interferometer. The set of such orthogonally invariant
states is convex and is therefore characterized by its extremal
points, namely the states
\begin{displaymath}
  \sigma_k^{(n)} = \frac{1}{a_k^{n}} \sum_{\stackrel{k_1 \cdots k_n}{\mathrm {s.t.}\; \sum_i k_i = k } }
  | k_1 \cdots k_n \rangle \langle k_1 \cdots k_n|
\end{displaymath}
where $|k_1 \cdots k_n\rangle$ is the $n$-mode Fock state with $k_i$
photons in mode $i$ and $a_k^{n} = {n+k-1 \choose n-1}$.

Physically, these extremal states are (proportional to) the
projectors onto the different eigenspaces of the total number operator
$\hat{n} = \hat{n}_1 + \cdots \hat{n}_n$ labelled with the integer
parameter $k$, corresponding to the total number of photons
distributed over the $n$ modes. The normalization constant $a_k^{n}$
simply counts the number of ways of distributing $k$ photons into $n$
modes. These extremal states $\sigma_k^{(n)}$ form a discrete infinite
set of mixed states. Importantly, any pure eigenstate chosen in the
eigenspace corresponding to a given total photon number $k$ is
generally not orthogonally invariant; only the uniform mixture of them
fulfills this invariance (Schur's lemma), which is why the extremal
states $\sigma_k^{(n)}$ are mixed for $n>1$.  Finally, any state $\rho$
that is invariant under orthogonal transformations in phase space can be written as
\begin{displaymath}
  \rho = \sum_{k=0}^{\infty} c_k \sigma_k^{(n)}
\end{displaymath}
where the weights $c_k$ satisfy $0 \leq c_k \leq 1$ and $\sum_k c_k
=1$.

\subsection{A Gaussian quantum de Finetti theorem for rotation-invariant states}
Let us now introduce a classical de Finetti theorem for continuous
variables. An infinite sequence of real-valued random variables $X_1,
\cdots, X_n \cdots$ is called {\em orthogonally invariant} if, for
every $n$, the probability distribution of $X_1,\cdots,X_n$ is invariant
under all orthogonal transformations of $\mathbb{R}^n$. It was proven in
\cite{Sch38,Fre62} that orthogonally invariant distributions are
exactly mixtures of i.i.d. normal distributions.

This result holds only approximately for finite sequences: if
$X_1,\cdots,X_n$ is invariant under orthogonal transformations of
$\mathbb{R}^n$, then there exists a mixture of i.i.d. normals such
that its variation distance to the marginal law of the first $k$
coordinates of $X_1,\cdots,X_n$ is bounded by $O(k/n)$ for $k\ll n$
\cite{DF87}. This cannot be directly applied to quantum systems, however,
since Wigner functions are not necessary legitimate probability
distributions (they can be negative). Here, we prove that this generalization
is nevertheless correct in the asymptotic regime.  In particular, 
we prove that an orthogonally invariant state tends to a mixture 
of multimode thermal states, which are products of $n$ thermal states
with the same mean photon number.
 
Let us consider an $n$-mode state $\rho$ which is orthogonally
invariant in phase space. For any $N>n$, $\rho$ is the partial trace
over $(N-n)$ modes of an $N$-mode orthogonally invariant state
$\rho^{(N)}$. As stated above, $\rho^{(N)}$ is a convex mixture of the
states $\sigma_k^N$. Therefore, it is enough to prove that the trace
over $(N-n)$ modes of $\sigma_k^{N}$ becomes asymptotically close
(for the trace distance) of a multimode thermal state as $N$ tends to
infinity.  Since the state of interest $\tr_{N-n} \; \sigma_k^{N}$ 
as well as the ``target'' $n$-mode thermal state $\rho_{\mathrm{th}}^n(x)$
with $k/n$ photons per mode are orthogonally invariant, they can both
be written as mixtures of $\sigma_l^{n}$'s,
\begin{displaymath}
  \tr_{N-n}\;\sigma_k^N = \sum_{l=0}^p f(l) \sigma_l^n \quad \qquad \quad
  \rho_{\mathrm{th}}^n =  \sum_{l=0}^{\infty} g(l) \sigma_l^n
\end{displaymath}
with
\begin{displaymath}
  f(l)=\frac{a_l^n \; a_{k-l}^{N-n}}{a_k^N} \quad \qquad \quad
  g(l)=a_l^n \frac{(k/N)^l}{(1+(k/N))^{n+l}}.
\end{displaymath}
Note that $f$ and $g$ also depend on $k$, $n$, and $N$, 
but we do not mention these parameters explicitly in order to simplify the notations.
The trace distance between the two states is given by the variation
distance between the two classical probability distributions $f$ and
$g$
\begin{displaymath}
  ||\tr_{N-n}\;\sigma_k^N - \rho_{\mathrm{th}}^n||_1 =  \sum_{l=0}^{\infty}|f(l)-g(l)|.
\end{displaymath}
The second member can be bounded from above as
\begin{eqnarray*}
  \sum_{l=0}^{\infty}|f(l)-g(l)|&=& \sum_{l=0}^{\infty} \left| \frac{f(l)}{g(l)}-1 \right| g(l) = 2\sum_{l=0}^{\infty}\left(\frac{f(l)}{g(l)}-1 \right)^+ g(l)\\
  &\leq& 2 \left(\sup_{l} \frac{f(l)}{g(l)} -1\right)
\end{eqnarray*}
where the last inequality follows from the triangle inequality, and $(x)^+$ stands for $x$ if $x \ge 0$
and 0 if $x<0$.  Let us introduce the notation
\begin{displaymath}
  h(l) \equiv  \frac{f(l)}{g(l)}=\frac{a_{k-l}^{N-n}}{a_k^N} \times \frac{(1+(k/N))^{n+l}}{(k/N)^l}.
\end{displaymath}
The rest of the proof consists in approximating $\sup h$ in the asymptotic regime.  
This is done by using the asymptotic approximation of $a_{xn}^{yn}$
with $n \to \infty$ resulting from Stirling's formula, namely
\begin{displaymath}
  a_{xn}^{yn} \sim \sqrt{\frac{1+y/x}{nx}} \; 2^{y \, n \, G(x/y)}
\end{displaymath}
where $G(z) = (z+1) \log_2(z+1) - z \log_2(z)$ is the von Neumann
entropy of a thermal state with $z$ photons.  Let us
introduce the reduced variables $x= k/N$, $y=n/N$, $z=l/N$ and
$t=(1-y)/(x-z)$. We can approximate the function of interest $h(l)$ as
\begin{displaymath}
  h(zN)= \frac{a_{N(x-z)}^{Nt(x-z)}}{a_{xN}^N} \frac{(1+x)^{N(y+z)}}{x^{Nz}} \sim A 2^{NB},
\end{displaymath}
where
\begin{displaymath}
  A = \sqrt{\frac{x(1+t)}{(x-z)(1+1/x)}} =  \sqrt{\frac{tx(1+t)}{(1-y)(1+1/x)}} 
\end{displaymath}
and
\begin{displaymath}
  B = (1-y) g(1/t) - g(x) +(y+z) \log(1+x) -z \log(x).
\end{displaymath}
Deriving $B$ with respect to $z$, one has
\begin{displaymath}
  \frac{\partial B}{\partial z} =- \log (1+t) + \log(1+1/x).
\end{displaymath}
Therefore, $B$ is extremal for $t =1/x$, that is $z=x y $, giving $B
\leq 0$.  As a result, one has
\begin{displaymath}
  \sup_l h(l) = \sup_z h(zN) \sim \frac{1}{\sqrt{1-y}} \sim 1 + \frac{n}{2N} \quad \mathrm{for} \quad n \ll N.
\end{displaymath}
Hence, $||\tr_{N-n}\;\sigma_k^N - \rho_{\mathrm{th}}^n||_1 \rightarrow
0$ for $N \rightarrow \infty$, which proves the quantum
continuous-variable version of the de Finetti theorem 
for orthogonally invariant states in the asymptotic regime.

\subsection{Bipartite case}

So far, we only discussed single-partite orthogonally-invariant
states. Obviously, in order to use this approach to the study of QKD
security, one needs a bipartite generalization. Let us consider the
case of a $2n$-mode bipartite state $\rho_{AB}$, meaning that Alice
and Bob each have $n$ modes.  Such a state $\rho_{AB}$ is termed 
{\em  invariant under conjugate orthogonal transformations} in phase space if, for any
Gaussian unitary operation $U$ corresponding to a real symplectic
orthogonal transformation in Alice's $2n$-dimensional phase space,
it satisfies
\begin{displaymath}
  U \otimes U^{*} \; \rho_{AB} \; U^{\dagger} \otimes U^{T} = \rho_{AB} 
\end{displaymath}
where $U^{*}$ is the Gaussian unitary operation corresponding 
to the conjugate orthogonal transformation in Bob's phase space. 
Physically, this invariance means
that $\rho_{AB}$ remains unchanged when Alice processes her $n$ modes
into any passive linear interferometer while Bob processes his $n$
modes into the passive linear interferometer effecting the conjugate
rotation in phase space.

Ideally, one should have a quantum de Finetti theorem for bipartite
orthogonally invariant states since this is the case which is directly
relevant for proving the security of continuous-variable QKD. 
The reason is that, following the arguments in Section \ref{symmetry}, 
Alice and Bob can indeed assume their bipartite state $\rho_{AB}$ to be invariant
under conjugate orthogonal transformations. 
Thus, a bipartite quantum de Finetti theorem would rigorously prove that $\rho_{AB}$ 
is ``close to'' a product of Gaussian states. Note, however, that an {\em exponential} version 
of the theorem would actually be required to address the security of continuous-variable QKD,
meaning that it is enough to trace over only an exponentially small number of modes 
in order to get a good approximation by a Gaussian state. Then, such a Gaussian state 
would actually be the product of $n$ i.i.d. Gaussian states, and the security
against collective attacks would therefore imply the security against
arbitrary attacks. 

Finding a bipartite version of this quantum de Finetti theorem is the subject
of further work. Although we do not have a rigorous proof yet, the fact that
a bipartite version of the theorem holds is very likely. In particular, 
both partial traces $\rho_A = \tr_B \,\rho_{AB}$ and $\rho_B = \tr_A \,\rho_{AB}$ 
are single-partite orthogonally-invariant states, for which the theorem applies.
Hence, locally, we already know that a state $\rho_{AB}$ that is
invariant under conjugate orthogonal transformations in phase space 
becomes asymptotically Gaussian. One only needs to prove that the correlations 
between Alice and Bob also behave according to the bipartite version of the theorem.

\section{Conclusion and perspectives}
\label{conclusion}

We have discussed the role of symmetries in the security analysis of QKD,
and introduced a new symmetry that is especially suited to continuous-variables schemes. 
This symmetry, which can be spelled out in the phase space representation, encompasses 
the usual symmetry under permutations in state space that have been considered so far 
in the context of discrete-variable QKD. We then derived an asymptotic quantum de Finetti theorem 
for orthogonally-invariant states in phase space, and showed that Gaussian states play
a role similar to that of i.i.d. states in the usual de Finetti theorem. More precisely,
any orthogonally-invariant state can be shown to be asymptotically close to a mixture 
of product Gaussian (thermal) states. This first application of a symmetry in phase space 
to the QKD security analysis seems very promising as Gaussian states have been known
to play a fundamental role in the analysis of continuous-variable QKD.



The perspectives of this work towards proving the unconditional 
security of continuous-variable QKD are twofold. A first approach would be 
to study the generalization of our (asymptotic) continuous-variable quantum de Finetti
theorem in phase space to the bipartite scenario, and then investigate whether 
an exponential version can be derived. The second option would be to see 
if the techniques recently introduced in \cite{CKR09} 
can be generalized to continuous variables.

\section*{Ackowledgements}
A.L. thanks Renato Renner and Johan {\AA}berg for fruitful discussions.
The authors acknowledge financial support of the European Union under project SECOQC
(IST-2002-506813), of Agence Nationale de la Recherche under
projects PROSPIQ (ANR-06-NANO-041-05) and SEQURE (ANR-07-SESU-011-01),
and of the Brussels-Capital Region under projects CRYPTASC and Prospective Research for Brussels.

\section*{References}

\end{document}